\documentclass[a4paper,12pt]{amsart}
\usepackage{amssymb,amsmath,mathrsfs}
\usepackage[usenames,dvipsnames]{xcolor}

\usepackage[stable]{footmisc}

\usepackage{hyperref}
\usepackage{tensor}
\usepackage{graphicx}
\usepackage{revsymb}
\usepackage[left=1in,top=1in,right=1in,bottom=1in,headheight=0.8in,foot=0.8in]{geometry}
\usepackage[nodisplayskipstretch]{setspace} 
\usepackage{mdframed}
\usepackage{tikz}
\usepackage[stable]{footmisc}
\expandafter\def\expandafter\normalsize\expandafter{%
    \normalsize%
    \setlength\abovedisplayskip{12pt}%
    \setlength\belowdisplayskip{12pt}%
    \setlength\abovedisplayshortskip{12pt}%
    \setlength\belowdisplayshortskip{12pt}%
}

\renewcommand{\footnoterule}{
  \kern 8pt
  \hrule width 0.4\linewidth height 0.4pt
  \kern 8pt 
}

\hypersetup{
	colorlinks=true,         
	linkcolor=MidnightBlue,          
	citecolor=MidnightBlue,
	urlcolor=MidnightBlue            
 }
\usepackage{natbib}
\setcitestyle{aysep={}} 

\makeatother

\newcounter{wn}

\title[On the Quantum Theory of Molecules]{{\textnormal{On the Quantum Theory of Molecules: \\ Rigour, Idealization, and Uncertainty}}}
\usepackage{amsaddr}

\author{Nick Huggett}
\address{\vspace{-0.8pc}University of Illinois Chicago \\Chicago, Illinois, USA}
\email{\href{huggett@uic.edu}{huggett@uic.edu}}

\author{James Ladyman}
\address{\vspace{-0.8pc}University of Bristol\\Bristol, United Kingdom}
\email{\href{James.Ladyman@bristol.ac.uk}{James.Ladyman@bristol.ac.uk}}

\author{Karim P. Y. Th\'ebault}
\address{\vspace{-0.8pc}University of Bristol\\Bristol, United Kingdom}
\email{\href{mailto:karim.thebault@bristol.ac.uk}{karim.thebault@bristol.ac.uk}}

\date{\today}

\begin{document}

\setstretch{1.2}

\begin{abstract}
Philosophers have claimed that: (a) Born-Oppenheimer approximation methods for solving molecular Schrödinger equations violate the Heisenberg uncertainty relations; therefore, (b) `quantum chemistry' is not fully quantum; and (c) therefore chemistry does not reduce to physics. This paper analyses the reasoning behind Born-Oppenheimer methods and shows that they are internally consistent and fully quantum mechanical, contrary to (a)-(c). Our analysis addresses important issues of mathematical rigour, physical idealization, reduction, and classicality in the quantum theory of molecules, and we propose an agenda for the philosophy of quantum chemistry more grounded in scientific practice.
\end{abstract}
\maketitle
\tableofcontents
 
\setstretch{1.5}

\section{Introduction}

Quantum chemistry is the use of quantum mechanics (and quantum field theory) to model molecules and their dynamics, with the aim of explaining and predicting their chemical properties and reactions.\footnote{An earlier draft of this paper appeared in preprint archives. In addition to clarifying our arguments, we revised our discussion of the work of Woolley and Sutcliffe in response to feedback from Olimpia Lombardi, Thierry Jecko, and Guy Woolley (see also \cite{lombardi2025talkingdiscussbornoppenheimerapproximation}).} Its status is important to questions about the place of physics among the sciences: prima facie, quantum chemistry is an example of the success of reductionism, exemplifying the fundamentality of physics with respect to chemistry and other natural sciences.\footnote{There are of course different notions of both reduction and fundamentality, and both epistemic and ontic versions of each.} However, the received view in the philosophy of quantum chemistry is that attention to models and scientific practice reveals not only that quantum chemistry does not reduce to quantum physics, but that the two are explicitly in conflict. One principal argument for this view is based upon alleged \textit{non-quantum} features of the `Born-Oppenheimer approximation' (\textbf{BO}), which is the most commonly used model of molecular dynamics in quantum chemistry. It has been claimed in the philosophy literature that \textbf{BO} violates the Heisenberg uncertainty principle, and so is in conflict with quantum theory.\footnote{This claim has been developed in most detail  by  Lombardi and coauthors \citep*{Lombardi2010,gonzalez:2019,fortin:2021,lombardi:2023}. See \cite{woolley:1977,woolley:1978,claverie:1980} for the original quantum chemistry discussion (which makes weaker claims). See also \cite{accorinti:2022}. Hendry \citep{hendry:1998,hendry:2006,hendry:2010,hendry:2010b, hendry:2017} while taking an anti-reductionist stance, does not claim that \textbf{BO} violates the uncertainty relations.  Our opposing analysis is  in the same spirit as \cite{ramsey:1997,scerri:2012,hettema:2017,franklin:2020,seifert:2020,seifert:2022,scerri2025born}.}

For instance, in an influential essay supporting anti-reductionism concerning the chemistry–physics relation (citing a talk by Olimpia Lombardi), Hasok Chang asserts:
\begin{quote}
...the typical method of quantum-mechanical treatment of molecules begins with the Born–Oppenheimer approximation, which separates out the nuclear wavefunction from the electronic wavefunction [...] Additionally, it is assumed that the nuclei have fixed positions in space. In this ``clamping-down'' approximation, the atomic nuclei are treated essentially as classical particles; as Olimpia Lombardi points out, this picture is non-quantum in a very fundamental way as the simultaneous assignment of fixed positions and fixed momenta (namely, zero) to them violates the Heisenberg uncertainty principle. But without such classical scene-setting, the quantum calculations are quite impossible. \cite[p. 198]{chang:2015}
\end{quote}
\textit{In our terms}, Chang claims that \textbf{BO} involves both the separation of wavefunctions (which is what \textit{he} calls `the Born-Oppenheimer approximation') and a `clamping-down' approximation, and that the latter violates the Heisenberg uncertainty principle, so \textbf{BO} as a whole is not fully quantum. Note that the claim is of `violation' of the principle rather than a mere conceptual or practical conflict. \cite{lombardi:2023} puts it even more strongly. When endorsing Torretti's (\citeyear{torretti:2000}) view that bringing together theories without worrying about their incompatibility, seemingly `outrageously' (p. 119), is nonetheless scientifically legitimate on pragmatic grounds, she notes, \textbf{BO}:
\begin{quote}
as used in the context of quantum chemistry, is a vivid example of how scientists ``outrageously'' appeal to incompatible theories in their practice. In this case, quantum chemical models of molecules are obtained by combining classical mechanics to describe the nuclei and quantum mechanics to account for the motion of the electrons. (p. 115)
\end{quote}
Lombardi and Chang's idea that \textbf{BO} violates Heisenberg uncertainty grounds more general anti-reductionist claims \citep{accorinti:2022,cartwright:2022}. Cartwright argues that the alleged violation implies not only that chemistry fails to reduce to physics, but that the two are inconsistent.\footnote{This case is discussed as part of Cartwright's general critique of reductionism and other forms of physicalism (this chapter of her book is entitled  `dethroning the queen').}  After quoting the passage from Chang above (without reference to Lombardi et al.), Cartwright says: 
\begin{quote}
 This approximation treats the atomic nucleus as a classical particle. But this fundamentally violates quantum mechanics which, following the Heisenberg uncertainty principle, maintains that we cannot have a simultaneous assignment of fixed positions and fixed momenta. The approximations that provide the reduction violate the very theory that the chemistry is being reduced to [...] the success of quantum chemistry relies fundamentally on assumptions that belong to classical chemistry \cite[pp. 106-7]{cartwright:2022}
\end{quote}
Note the claim is that the approximation `fundamentally violates quantum mechanics'.

To our knowledge, the idea that \textbf{BO} might violate the Heisenberg uncertainty principle has never been discussed in the considerable mathematical physics literature that has been devoted to the study and development of \textbf{BO}. Indeed, from a mathematical perspective, a genuine violation of the Heisenberg uncertainty principle could only occur in a context beyond the Hilbert space representation of states and observables that is core to \textbf{BO}.\footnote{\label{mixed} An example of such a representation is found in mixed classical-quantum models which simultaneously feature representations of classical and quantum states and dynamics \citep{tully:1991,crespo2018recent}. In this context, there is the possibility of failure of positivity of the density matrix in the quantum part of the model which would imply violation of the Cauchy-Schwartz inequality which, in turn, implies violation of the Heisenberg uncertainty type relations \citep{bondarenko:2023,gay:2023}. Significantly, such a feature is understood by the scientists themselves as a pathological feature of the models rather than a putative representation of Heisenberg uncertainty violation or, moreover, a failure of reduction. Indeed, work on the topic takes density matrix positivity to be a precondition of physically consistent mixed classical-quantum dynamics.} Thus, according to the basic structure of the model, there is simply no possibility for the uncertainty principle to be violated. This notwithstanding, versions of the erroneous claim that it does have been expressed (largely informally) with the quantum chemistry literature.\footnote{See for example, \cite[pp. 100-1]{villaveces:1990} and \cite[p. 1760]{lang:2024}. Note, however, that the \textit{explicit claim} that the \text{BO} violates the Heisenberg uncertainty principle is never expressed in the notable discussions of Sutcliffe and Woolley cited earlier. \cite[p. 281-2]{woolley:2022} might suggest that such an issue would occur in the $\kappa=0$ limit of the \textbf{PBO} to be discussed in \S\ref{history}. However, that is a completely different claim.} 

The first aim of this paper is to demonstrate unambiguously that \textit{\textbf{BO} does not violate the Heisenberg uncertainty principle.}\footnote{For related critical remarks see \cite{scerri2025born}. We give a more thorough analysis, but the latter paper addresses related challenges beyond the scope of this paper.} We show this by analysing a textbook-style presentation of the modern \textbf{BO}, given in \S\ref{textbook}-\ref{sec:derivations}, which we contrast with historical versions (\S\ref{history}). Of course, there is more to chemistry than can be shown using \textbf{BO}, so our demonstration does not suffice to show its reduction to physics. However, it does clarify the relation, and refute one powerful anti-reductionist argument. 

The second aim of this paper is to provide an account of the idealizations operative within the modern version of the \textbf{BO}. We first analyse the structure of \textbf{BO} as an idealized quantum mechanical model. We then provide  justification for the key idealization within the model based upon the stability of the salient model-based inferences regarding the behaviour of molecules under de-idealization. We present the core details of this analysis in \S\ref{Idealization} and a more detailed justification in \S\ref{textbook}.  

The third aim of this paper is to address concerns about the rigour of the textbook presentation of \textbf{BO}: in particular, subtle issues regarding the justification of the mathematical idealizations involved in modern formalizations and the assumptions needed in the derivation of a discrete spectrum. First,  \cite{sutcliffe2012quantum, Sut+Woo13} articulate convincing criticism of \cite{Born:1954}, regarding the spectrum of the `electronic Hamiltonian' and the interpretation of \textbf{BO} as a perturbative analysis. In \S\ref{Jecko} we summarise this issue and argue that it does not apply to our presentation of \textbf{BO}. We also highlight the analysis of \cite{Jecko:2014}, whose stated goal was to temper Sutcliffe and Woolley's `pessimism' about the tenability of \textbf{BO} as a quantum approximation, by outlining how mathematical physics has provided a rigorous approach to \textbf{BO}.

The fourth and final aim of this paper is to use the analysis of \textbf{BO} to open up wider questions concerning the role of reduction and rigour in quantum chemistry. \S\ref{BBO} provides a prospectus for future philosophical work on the foundations of quantum chemistry that is informed by scientific practice, as all parties agree that it should be. We argue that such philosophical work should be disentangled from the unwarranted mobilization of quantum chemistry against the fundamentality of physics. We show that attending instead to the conceptual, formal and methodological questions which scientists themselves ask raises a range of issues and open questions relating to the various types of semi-classical modelling, the role of persistent environmental interactions, and the problem of isolating distinctively `chemical' modes of quantum modelling practice.

\section{Idealized Quantum Models of Molecules }
\label{Idealization}

When quantum chemists and physicists talk about \textbf{BO} they refer to an approach to solving the quantum mechanical equations for a molecule, based on an \textit{idealized quantum model} of a molecule that builds on -- but modifies and extends --  pioneering work by Born and Oppenheimer in 1927. Before going into details, it will be helpful to give an overview of \textbf{BO}, its idealizations and their rationale.

The basic idea to use to the high ratio between the electron and nuclear masses to produce trial solutions to  a molecular Schr\"odinger equation: we consider the time-independent equation, but the method extends to dynamical problems as well. There are two distinct aspects of \textbf{BO}: a \textit{separation ansatz} and an \textit{adiabatic approximation}. The ansatz is that the molecular wavefunction, $\Psi(x_1,x_2)$, is approximated by the product of a function of nuclei positions, $\theta_a(x_1)$, and a (specified) function of nuclei and electron positions, $ \psi_a(x_1,x_2)$, so that we have: $\Psi(x_1,x_2) =  \theta_a(x_1)\psi_a(x_1,x_2)$. The approximation is that the rate of change of the $\psi_a(x_1,x_2)$ with respect to the nuclear position is approximately zero; this condition provides an equation for $\theta_a(x_1)$.

The arguments of  \cite{Lombardi2010,gonzalez:2019,fortin:2021} (and perhaps also of \cite{chang:2015}, which endorses Lombardi's analysis in the passage above but does not otherwise elaborate) involve a  misrepresentation of how \textbf{BO} is justified in quantum chemical practice. This is not done, as they suggest, by idealizing the nuclei as classical, but by a quantum theoretical idealization about energy levels, justified by its stability under de-idealization (as in many other cases). The \textit{intuitive} idea is that the total kinetic energy of the nuclei is small compared to the sum of the potential energy of the molecule and the total kinetic energy of the electrons. This is because a nucleon is far heavier than an electron, and so typically moves far more slowly, and kinetic energy scales as $mv^2$.

\begin{figure}
    \centering

\begin{tikzpicture}

\draw[->] (0,0) -- (2.25,0) node[anchor=north]{$\mathrm{x}$} -- (5,0) node[anchor=west]{$x_1$};
\draw[->] (0,0) -- (0,4) node[anchor=south]{$E$};
\draw (0,1) .. controls (2.25,0) .. (5,1.5) node[anchor=west]{$\lambda_1(x_1)$};
\draw (0,3.5) .. controls (1.75,1.5) .. (5,2.5) node[anchor=west]{$\lambda_2(x_1)$};
\draw (0,2) .. controls (2.25,.75) .. (5,3.2) node[anchor=west]{$\lambda_3(x_1)$};

\end{tikzpicture}

    \caption{Eigenvalues $\lambda_n$ of the clamped Hamiltonian, as (hypothetical) functions of the heavy, nuclear degrees of freedom $x_1$. In the region around $x_1=\mathrm{x}$ the first three electronic energy levels can be seen to be widely separated: specifically, by far more than the kinetic energy of the nuclei. This is the condition for stable molecules, and for \textbf{BO}.}
    \label{fig:BO}

\end{figure}
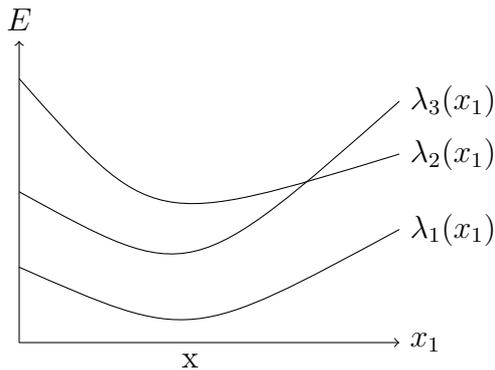

\textit{Mathematically}, $\psi_a(x_1,x_2)$ is a \textit{family} of electron ($x_2$) wavefunctions, one for each fixed nuclei configuration ($x_1$). Each function in the family is an eigenstate of a so-called `clamped Hamiltonian', the sum of terms for the kinetic energy and potential energy of the electrons, given that nuclei configuration.\footnote{`So called' to emphasize that the proposition we formulate as \textit{Clamped} (see below) is not assumed or implied by the name of this mathematical object.} The corresponding family of energy eigenvalues $\lambda_a(x_1)$ picks out a `potential energy surface' (PES) as illustrated in Figure \ref{fig:BO}. (\S\ref{textbook} explains in detail why it would be a confusion to think that these merely formal facts in any way amount to idealizing the nuclei as classical.)  $\theta_a(x_1)$ is, formally speaking again, a wavefunction for the nuclei in this potential. The molecular energy is approximately the sum of $\lambda_a(x_1)$ and the nuclear kinetic energy, and because of the energy difference, close to the former (for any relevant $x_1$). So if the gaps between the $\lambda_a(x_1)$ are large, only one $\psi_a(x_1,x_2)$ is relevant to $\Psi(x_1,x_2)$ -- superposing with other eigenstates of the clamped Hamiltonian shifts the energy too far. Figure \ref{fig:BO} shows the crucial representative features of the PESs. It makes clear that energy gaps do not exist for all values of $x_1$. It also illustrates that the gap can exist for a given range of values, about a minimum.

One expects that the nuclei of a stable molecule are restricted to a region around such a point; the nuclear wavefunction effectively vanishes outside. Thus the more specific assumption made in \textbf{BO} is:
\begin{quote}
 \textit{Heavy}: in a stable molecule the nuclei are approximately localized, in a \textit{quantum} state in which their kinetic energy is much smaller than the electron kinetic energy (though not zero).
\end{quote}
In \S\ref{textbook}, once some of the necessary formalism has been developed, we show how (a precisification of) this assumption justifies the separation ansatz and adiabatic approximation, and hence \textbf{BO}, in quantum mechanical terms. We emphasize that this justification does \textit{not} involve taking  nuclear masses to infinity: indeed, such a limit is not generally well-defined \citep{PhysRevA.59.4270, sutcliffe2012quantum}. Manifestly, \textit{Heavy} does not assert the existence of such a limit.

The idealized model is one in which \textit{Heavy} holds for \textit{any} nuclear configuration, $x_1$, corresponding to a stable molecule. That is, the energy gaps exist for \textit{all} values of $x_1$. In \textbf{BO}, generically, they do not: there are regions like that in Figure \ref{fig:BO} in which the levels cross, and the idealization fails. However, on one standard account of idealization, inferences based upon an idealized model are justified to the extent that they are stable under relaxation of the idealization \citep{McMullin1985,landsman:2013,earman:2004}.\footnote{For further relevant discussion of idealization and approximation see \cite{norton:2012,bokulich:2017,palacios:2024}.} Suppose, then, that \textit{Heavy} is valid in the region in which the nuclei are effectively restricted. Then, to the extent that the support of the full molecular wavefunction lies in the region in which \textit{Heavy} holds, we can expect the wavefunction to remain approximately separable -- for stability under de-idealization.\footnote{\label{LBBO}The adiabatic approximation can be expected to start to break down in various circumstances. One obvious case is when the nuclei are light, as with hydrogen. \cite{yang:2023} note: ``considering the high mobility of light hydrogen atoms, the non-adiabatic coupling of different electronic states beyond the Born-Oppenheimer approximation is expected to be prominent'' (p. 2). Such coupling gives rise to conical intersections between PESs that are used to understand reaction pathways \citep{baer:2006}. Much of current work in quantum chemistry goes beyond \textbf{BO} by considering the interactions between electronic and nuclear vibrational motion which leads to the coupling of different energy states of molecules \citep{yarkony:2012}. See also \cite{sibaev:2020,agostini:2022} } We return to this point in \S\ref{textbook}. 

The idealized model includes a false assumption regarding the range of validity of \emph{Heavy}, but does not involve any assertions inconsistent with quantum theory. By contrast, consider an alternative idealizing assumption of the form: 
\begin{quote}
    \emph{Clamped}: molecular nuclei have fixed definite positions and zero kinetic energy.\footnote{E.g., the `clamped nuclei approximation' of \cite[pp.41-3]{gonzalez:2019}.}
\end{quote}
An idealized model of which \emph{Clamped} is true would be one in which molecular nuclei have classical positions and momenta (namely zero), in conflict with the Heisenberg uncertainty principle, as claimed in the quoted passages. In such a model it would be plausible to argue that a classical modelling procedure is required to apply the approximations involved (even if it is only one half of an `outrageous' classical-quantum combination ultimately delivering a fully quantum solution). However, the authors mentioned above do not show that such an idealization is part of the form of \textbf{BO} which we discuss below (or indeed its original form). Nor does use of the \textit{family} of clamped Hamiltonians assume \textit{Clamped}, which effectively selects \textit{one} of them. Moreover, \textit{Heavy} is a much weaker assumption than \textit{Clamped} and consistent with the nucleus being fully quantum. Indeed, (a precisification of) \emph{Heavy} justifies the separation ansatz and adiabatic approximation, and hence justifies \textbf{BO}, in quantum theoretic terms. This demonstration involves neither of the stronger (arguably false) assumptions of a mass limit or \textit{Clamped}. All these things are shown in \S\ref{textbook}-\ref{sec:derivations}, the core of our argument concerning \textbf{BO}.

To summarize, our argument regarding idealizations and \textbf{BO} is as follows:
\bigskip
\begin{mdframed}
\centerline{Idealization and \textbf{BO}}
\begin{enumerate}
    \item[I1.] Idealized models are related to less idealized models via approximation relations.
    \item[I2.] Inferences based upon idealized models are justified if the features of the idealized model that ground the relevant inference are stable under de-idealization
\item[I3.] The fundamental idealization of \textbf{BO} is \textit{Heavy}. Inferences about the behaviour of molecules in the relevant regime are indeed stable under de-idealization 
\item [I4.] \textbf{BO} does not involve the idealization \textit{Clamped}.   
\item [I5.] The conflict with the Heisenberg uncertainly relation (and stability under de-idealization) of \textit{Clamped} is irrelevant to the use of models based on \textbf{BO}. 
\end{enumerate}
\end{mdframed}
\bigskip

\section{The Born-Oppenheimer Approximation}\label{sec:BOapprox}
\subsection{The Historical Treatment}
\label{history}

What we can call the `Perturbative Expansion Born-Oppenheimer approximation' (\textbf{PBO}) is the approach to the approximate solution of the (time-independent) Schr\"{o}dinger equation for stable molecules pioneered by Born and Oppenheimer in their 1927 paper \textit{Zur Quantentheorie der Molekeln} (`On the Quantum Theory of Molecules').\footnote{See \cite{BOTrans}, \cite{sutcliffe2012quantum} and \cite{scerri2025born}.} The key physical feature upon which the \textbf{PBO} is built is that nuclei are very much heavier than electrons. With this in mind Born and Oppenheimer introduce a small parameter, $\kappa = \Big{(} \frac{m}{M}\Big{)}^\frac{1}{4}$, where $m$ is the electron mass and $M$ the nucleon mass; the ratio $m/M$ is roughly $1/2000$ so $\kappa\approx0.15$. The crucial idea is to treat the nuclear kinetic energy as a perturbation of the energy, expanded in powers of $\kappa$. The original \textbf{PBO} is indicated by the authors to be valid from zeroth to fourth order in $\kappa$ with nuclear vibrational energy corresponding to terms of second order and the rotational energy to fourth order in the energy. Coupling effects among electronic states appear beyond fourth order in $\kappa$.

The principal achievement of \textbf{PBO} is to show that one can use the Hamiltonian for electrons in the Coulomb field of a fixed nuclear configuration to construct a family of electronic wavefunctions. These wavefunctions can then be used to calculate approximate eigenvalues for the full molecular Hamiltonian on the assumption that the nuclear motion is confined to a small vicinity of a privileged equilibrium configuration. Up to order $\kappa^4$, approximate wavefunctions can then be written as products of  `electronic' wavefunction and `nuclear' wavefunctions  \citep[p.3]{sutcliffe2012quantum} (the physical significance of these objects is discussed in more detail in following sections).

The seeds of later confusion were sown in the description that Born and Oppenheimer provide of the zeroth order equations. In particular, Part II of the paper, titled `Electronic Motion for Stationary Nuclei' states that `if one sets $\kappa=0$ one obtains a differential equation in the [electron position variables] alone, the [nuclear position variables] appearing as parameters'. The solution of such an equation is then indicated to `represent the electronic motion for stationary nuclei'. As noted by \cite{sutcliffe2012quantum},  `it is perhaps to this statement that the idea of an electronic Hamiltonian with fixed nuclei as arising by letting the nuclear masses increase without limit, can be traced' (p. 2).

Yet, the representational content of a perturbative model should not be conflated with its zeroth order terms, on pain of misunderstanding their ubiquitous use as scientific models. \textbf{PBO} does \textit{not} represent the nuclei as fixed; rather it organizes an expansion in which the \textit{only} leading, zeroth order, term has that character. It expands around a \textit{fictitious} system in which the nuclei are clamped. These are entirely different modelling strategies. (Analogously, one should not conflate a purely Newtonian model with a perturbative expansion involving the Newtonian model plus relativistic corrections.) The result of \textbf{PBO} is, of course, (in principle) a wavefunction that satisfies the uncertainty relations for both electrons and nuclei. Thus, the idealizing assumption \textit{Clamped} does not form part of \textbf{PBO} and the idealization problem does not occur.\footnote{These remarks are not intended to address Sutcliffe and Woolley, rather we will turn to their work in \S\ref{Jecko}.}

\textbf{PBO} was made redundant by later work of  \cite{born:1951} and \cite{Born:1954}, and it is that approach that is today referred to in chemistry literature as defining the ``Born-Oppenheimer approximation'' and to which we now turn.

\subsection{A Textbook-style Presentation: preliminaries}
\label{textbook}

This section and the next are a treatment of the current  \textbf{BO}, drawing in parts upon  \cite{messiah:1962} and \cite{Jecko:2014} (but with no serious attempt at historical reconstruction). Here `textbook-style' indicates both accessibility and an emphasis on heuristics over rigour at certain points. In particular, the pedagogical aim of a textbook may be to \textit{sketch the overall logic} of a calculational method in order to emphasise certain principles, so that the method is memorable and portable to related problems in scientific practice. In the process certain idealizations have to be made, which should be justified by a more rigorous de-idealization in the background. Our treatment similarly describes the overall logic of \textbf{BO} in a way to make clear its quantum nature; it makes certain idealizations, (some of) which are flagged below and addressed in detail in \S\ref{Jecko}. In presenting at this level of informality, we follow our interlocutors, while aiming to deepen the level of mathematical precision in philosophical discussions of \textbf{BO}.

Suppose a system is comprised of two parts, with canonical variables $x_1 \in \mathbb{R}^m$ and $x_2 \in \mathbb{R}^n$. In the usual way, the state of the system in the $x$-representation -- the wavefunction -- is $\Psi(x_1,x_2)\in L^2(\mathbb{R}^m\times\mathbb{R}^n)$, and (setting $\hbar=1$) the canonically conjugate observables are $i\partial/\partial x_j$. The (non-relativistic) Hamiltonian is the sum of kinetic, $\hat T_j$, and interaction, $\hat W$, parts:
\begin{equation}\label{eq:HamTot}
    \hat H =\hat T_1 +\hat T_2 +\hat W,
\end{equation}
with $\hat W$ some function of the variables $x_j$, and $\hat T_j$ a power (or sum of powers) of the corresponding conjugate variables, $\partial^n/\partial x_j^n$. In the case of a molecule, the first two terms will be the kinetic energies of nuclei and electrons ( $\hat p_j^2/2m \propto -\partial^2/\partial x^2_j$), respectively, and the third, Coulombic potential energy.\footnote{One idealization has already been made; as discussed in \S\ref{Jecko}, a total kinetic energy due to the velocity of the molecule as a whole should be subtracted.}

We are interested in finding the eigenstates of $\hat H$, which cannot be done analytically, but requires approximation (at least in practice). So suppose further -- and this is the \emph{crucial assumption} -- that in a range of states of interest the kinetic energy of the $x_1$ subsystem is far smaller than both that of the $x_2$ part, and that of their interaction energy: $T_1 \ll T_2, W$ (where $O$ denotes the expected value of observable $\hat O$). What `range of states'? In the first place, those wavefunctions that only have (non-negligible) support in a finite range of values of $x_1$; the first system is effectively localized within that region. (Moreover, the states should be below some maximum energy level.)

The \textit{crucial assumption} (elaborated and sharpened below) is a mathematical \textit{ansatz}, a temporary supposition made to find solutions to an equation, which must then be inspected to verify that they really solve it: if so the supposition is vindicated; if not then it must be given up. For example, a familiar move when solving a differential equation for a function in two variables is to see whether it is the product of functions of only one variable, $F(x,y)=f(x)\cdot g(y)$. There are tests for such separability (which is \textit{not} the same as that in the \textbf{BO}), but they do not apply generally, and even when they do it may be simpler to try to solve the equation under the ansatz. If one is lucky then one finds such a function which demonstrably solves the equation, and one knows, rather than supposes, that there is a separable solution; if one is unlucky then of course the ansatz is rejected. An ansatz is then not intrinsically an approximation or idealization, but a guess, whose proof is in the pudding. However, one might make the ansatz that an approximate solution of certain kind exists: that a separable function approximates a true solution, say. Naturally, any such function would then have to be checked to see if it was an approximate solution, to see whether the guess was good. We stress that whether the ansatz postulates a feature of exact or approximate solutions, it never \textit{changes} those solutions; it just guesses what they are in fact like. All of this is a completely standard heuristic in mathematics, in science, and indeed in any reasoning, though it seems to have had little explicit discussion in philosophy.

One can seek motivation for making an ansatz, a reason to think that it will turn out to be true of some solution. But such \textit{motivation} should not be understood as its \textit{justification} -- if an ansatz is justified, it is wholly by the solutions it produces. In the present case the ansatz made to solve the molecular Schr\"odinger equation -- the \textit{crucial assumption} -- is motivated by the composition of a molecule, in which the nuclei comprise the first system, and the electrons the second. In typical states, because electrons are 2000 times lighter than nucleons, they are more easily set in motion, and a stable state is one in which almost all the kinetic energy is in the motion of the former; if it is a low energy state then the nuclei can only be displaced from a potential energy minimum by a small amount \citep[XVIII.12]{messiah:1962}. Notice that this motivation mixes physical and mathematical considerations: appealing to the physical features represented by elements of the formalism, and expectations about their physical behaviour. The justification, however, is purely mathematical: finding (approximate) solutions to the Schr\"odinger equation in which the guess is true. Moreover, as indicated in \S\ref{Idealization}, such confirmation also demonstrates stability under relaxation of the idealizations involved, again by showing that the (in practice approximate) solutions of idealized model remain approximate solutions of the (more) exact equations.

To see the role of the \textit{crucial assumption} in finding (approximate) eigenstates of $\hat H$, we first consider normalized solutions to the (time-independent) Schr\"odinger equation for the `clamped' Hamiltonian (in the $x$-representation) for a \textit{specified} value of $x_1$:
\begin{equation}
    \label{eq:RedHamTISE}
    \big(\hat T_2 + \hat W(x_1)\big)\psi_a(x_1;x_2) = \lambda_a(x_1)\psi_a(x_1;x_2),
\end{equation}
with $a=1,2,\dots$.\footnote{Part of the spectrum of $\hat T_2+\hat W(x_1)$ is continuous, so not all such states are normalizable, square-integrable functions. In this section we follow standard practice and ignore this complication; see \S\ref{Jecko}.} We cannot stress strongly enough that here we are thinking purely mathematically, about some formal properties of this equation, and are not taking the equation to represent -- even approximately -- anything physical, including a molecule. It it were to represent something, it could be a \textit{fictional} collection of electrons moving in the field of some fixed, or `clamped', charges. (The constants that appear in (\ref{eq:RedHamTISE}) are numerically equal to the masses and charges of molecular constituents.) But at present we are speaking formally, about the mathematics, not materially about any molecules. Indeed, as we are showing, at \textit{no} point in \textbf{BO} does a single clamped Hamiltonian represent the molecule under consideration.

Since different values for $x_1\in\mathbb{R}^m$ could be specified, there is a parameterized family of $x_2$ operators $\hat W(x_1)$: so there is not just one but infinitely many clamped Hamiltonian\textit{s}. In that case, the $\hat T_2 + \hat W(x_1)$ eigenvalues $\lambda_a(x_1)$ and eigenstates $\psi_a(x_1;x_2)$ are also parameterized families; hence the semi-colon. That is, formally speaking, the energy and state of the electronic subsystem vary for \textit{fixed} energy level $a$, as $x_1$ varies, the former variation tracing out a PES.

Understanding our philosophical claims about \textbf{BO} requires understanding some of the formal properties of this family. Because they are a set of eigenstates for a self-adjoint operator of the $x_2$  subsystem, for \textit{each}  $x_1=X\in\mathbb{R}^m$ the $\psi_a(X;x_2)$ form a complete orthonormal basis:$\int\psi_a^*\psi_b\text{d}x_2|_{x_1=X}=\delta_{ab}$.\footnote{Here and at many points in the following we make the important idealization that the spectrum is discrete. We discuss the removal of this assumption in \S\ref{Jecko}.} It follows mathematically (since in addition $\hat T_2+\hat W$ commutes with $\hat x_1$) that any $(x_1,x_2)$ wavefunction on $\mathbb{R}^m\times\mathbb{R}^n$ can be written:
\begin{equation}\label{eq:BOexpand1}
    \sum_a \theta_a(x_1)\psi_a(x_1,x_2).
\end{equation}
Note that this expansion is \textit{not} a sum of nuclear $\{\chi_a(x_1)\}$ and electron$\{\zeta_a(x_2)\}$ (tensor) product states:
\begin{equation}\label{eq:TPexpand}
    \sum_{a}\chi_a(x_1)\zeta_a(x_2).
\end{equation}
No, $\psi_a(x_1,x_2)$ represents a `direct integral', taking, for each $x_1\in\mathbb{R}^n$, an $x_2$ wavefunction $\psi_a(x_1;\cdot)$ satisfying (\ref{eq:RedHamTISE}) from a \textit{distinct copy} of the $L^2(\mathbb{R}^n)$ Hilbert space. Both (\ref{eq:BOexpand1}) and (\ref{eq:TPexpand}) give the general form of $L^2(\mathbb{R}^m\times\mathbb{R}^n)$ functions: the latter from the familiar properties of the tensor product; the former simply because specifying an $L^2(\mathbb{R}^n)$ function for each value of $x_1$ (in a suitably smooth way) specifies such a function. Inverting the picture, $\psi_a(X;x_2)$ is just the \textit{cross-section} of $\psi_a(x_1,x_2)$ at $x_1=X$.\footnote{Thus contrast (\ref{eq:BOexpand1}), with (\ref{eq:RedHamTISE}); the former expresses a function over $\mathbb{R}^m\times\mathbb{R}^n$, while the latter expresses a continuous infinity of equations for $x_2$ wavefunctions, one for each value of $x_1$. The use, in $\psi_a$, of a comma in former versus a semi-colon in the latter indicates just this difference.} So one cannot read (the terms in the sum) (\ref{eq:BOexpand1}) as describing separate $x_1$ and $x_2$ states: rather $\psi_a(x_1,x_2)$ is a wavefunction for \textit{both} parts (unlike an $x_2$ wavefunction $\psi_a(X;\cdot)$). Why expand such cross-sections in the eigenbasis of the clamped Hamiltonians rather than some other orthonormal basis? Because, as we shall soon see, in this basis one term in the (\ref{eq:BOexpand1}) dominates, and so we have (approximate) separability. This fact is the reason that we consider the clamped Hamiltonians, not because they themselves approximate (or represent!) anything; establishing it is the main point of the derivation in the next section.

To try to solve the family of equations of the form (\ref{eq:RedHamTISE}) one precisifies the \textit{crucial assumption}, and makes as an ansatz (exactly as discussed above):
\begin{quote}
    \emph{Heavy}: the \emph{gaps}, $|\lambda_n(x_1)-\lambda_m(x'_1)|$, between the $\lambda_a(x_1)$s are much greater than the \emph{values} of $T_1$, when compared for any $x_1$ and $x'_1$. 
\end{quote}
\textit{Heavy} is not merely an ansatz, it is also manifestly an idealization: inspection of Figure \ref{fig:BO} shows that different $\lambda_a(x_1)$ can cross, and that even when they don't it can be the case that $\lambda_a(x_1)\approx\lambda_b(x_1')$. However, one guesses that there are ranges of $x_1$ for which the condition holds, and that there are eigenstates of $\hat H$ whose support approximately lies in such a region. For a molecule, for instance, this amounts to the assumption that there are energy eigenstates in which the nuclei are sufficiently localized at the bottom of a potential energy well, which is quite reasonable for small excitation levels. Once again, this is a supposition about the mathematical -- and \textit{quantum} -- properties of some solutions of our equations, based on the physics. We have not changed the problem or its solutions in any way (except focussing on a subset), just helped ourselves find them, if indeed they are as we suppose. This guess of course needs to justified by checking that putative solutions to, for example, the molecular Schr\"odinger equation obtained by making it are indeed of this kind, even if they break \textit{Heavy} in the ways just described. Furthermore, verifying that putative solutions do indeed approximately solve the equation, despite the failure of \textit{Heavy} outside the relevant region, shows that it is stable under the necessary de-idealization -- as claimed in \S\ref{Idealization}.

By now, the reader may feel that a certain amount of hand waving is occurring. However, as explained, our goal here is not rigorously to \textit{prove} that the \textbf{BO} is valid under certain conditions, but to give an intuitive account of the mathematical and physical significance of the conditions. In particular, separability and adiabaticity follow quantum mechanically for solutions satisfying \textit{Heavy} (without taking a limit of the mass ratio), and these suffice for \textbf{BO} (without modeling the nuclei as clamped). The approximation has been subject to the more rigorous attentions of mathematical physicists, so the argument outlined here rests on solid mathematical ground (see \S\ref{Jecko} and \cite{Jecko:2014} for a review).

\subsection{A Textbook-style Presentation: the approximation}
\label{sec:derivations}

\emph{Heavy} has two important consequences whose derivation we now sketch.\footnote{The following arguments are presented without careful attention to the distinction between $x_1$-parameterized families of wavefunctions and operators on the one hand, and their direct integrals on the other. They are best read as equations relating the corresponding differential operators and functions in the position representation, which is indifferent to the distinction.} The arguments are straightforward, and in both cases the important point is that they follow from \textit{Heavy} alone (with no appeal to mass limits or \textit{Clamped}).

\textit{Derivation 1. Separability Ansatz from Heavy}.  There are eigenstates of $\hat H$ with the approximate form $\theta_a(x_1)\psi_a(x_1,x_2)$: the `Born-Oppenheimer ansatz'. To see this, first recall that it is a mathematical fact that any state can be written in the form (\ref{eq:BOexpand1}), so all we need to show is that eigenstates approximate a single term in the sum. So suppose (for reductio) that they don't; more specifically, to simplify things, suppose that $E$-valued energy eigenstate $\Psi(x_1,x_2)$ has non-negligible (suppose equal) contributions from two\footnote{Considering two terms is not innocuous as strictly the following depends on it; but it is however illustrative of the role energy gaps play in \textbf{BO} in general.} different (orthonormal)  $\theta_a(x_1)\psi_a(x_1,x_2)$, with energies $\lambda_m(x_1) < \lambda_n(x_1)$: 
\begin{eqnarray}\label{eq:whydiag}
\nonumber    \hat H\Psi &=& \hat H\frac{1}{\sqrt{2}}(\theta_m\psi_m+\theta_n\psi_n)
    = \frac{E}{\sqrt{2}}(\theta_m\psi_m+\theta_n\psi_n)\\
 &=& (\hat T_1+\hat T_2+\hat W)\frac{1}{\sqrt{2}}(\theta_m\psi_m+\theta_n\psi_n)\\ \nonumber
&=& \hat T_1\frac{1}{\sqrt{2}}(\theta_m\psi_m+\theta_n\psi_n) + \frac{\lambda_m}{\sqrt{2}}\theta_m\psi_m+\frac{\lambda_n}{\sqrt{2}}\theta_n\psi_n,
\end{eqnarray}
using (\ref{eq:HamTot}) and (\ref{eq:RedHamTISE}).\footnote{In the final step we use the fact that $\hat T_2$ contains only $x_2$ derivatives, while $\hat W$ is a function of $x_1$ and $x_2$, so both operators commute with $\theta_a(x_1)$; we also use the direct integral of the $\hat W(x_1)$, discussed in \S\ref{Jecko}. The argument does not depend on our simplifying assumption of equal, real amplitudes.}

The sum of the second two terms is a vector that fails to be parallel to $\Psi$ by a vector whose amplitude is the order of $(\lambda_n-\lambda_m)/\sqrt{2}$: for instance, one could either add $(\lambda_n-\lambda_m)\theta_m\psi_m/\sqrt{2}$ or subtract $(\lambda_n-\lambda_m)\theta_n\psi_n/\sqrt{2}$. That is to say, by (\ref{eq:whydiag}) -- namely the supposition that $\Psi$ is an eigenstate -- we have
\begin{equation}
    |\hat T_1\frac{1}{\sqrt{2}}(\theta_m\psi_m+\theta_n\psi_n)|\approx(\lambda_n-\lambda_m)/\sqrt{2}.
\end{equation}
But for any Hermitian operator and normalized vector, $|\hat O\phi|$ cannot exceed the greatest eigenvalue. So in this case, by \emph{Heavy}, 
\begin{equation}
    |\hat T_1\frac{1}{\sqrt{2}}(\theta_m\psi_m+\theta_n\psi_n)|\leq T_1^{max} \ll(\lambda_n-\lambda_m)/\sqrt{2},
\end{equation}
a manifest contradiction. Hence the supposition is false, and an eigenstate of total energy cannot be a sum of $\theta_a\psi_a$, but has the product form
\begin{equation}
    \label{eq:BOansatz}
    \hat H\Psi_a(x_1,x_2)\approx\hat H\theta_a(x_1)\psi_a(x_1,x_2) \approx E_a\theta_a(x_1)\psi_a(x_1,x_2).
\end{equation}
\begin{flushright}
$\square$
\end{flushright}
Put another way, such states (approximately) \emph{diagonalize} the total Hamiltonian: there are no cross-terms for such states with different values of $a$. (Note that in this approach, despite its historical name, \textit{separability} is not an independent ansatz, but rather the consequence of the \textit{Heavy} ansatz.) 

The derivation depends crucially on the choice of bases $\{\psi_a(x_1,x_2)\}$ for the expansion, so that \textit{Heavy} can be applied; indeed, in other bases it will not be the case that one term dominates, so the molecular wavefunction will not be even approximately separable. That is the \textit{significance of the clamped Hamiltonian}(s) in this explication of \textbf{BO}, not that any of them literally approximates or idealizes or replaces the true Hamiltonian. In this sense, clamping is entirely formal, not and does \textit{not} represent the nuclei of the target molecule as fixed. (The system of electrons moving in the Coulomb field of some fixed charges that it could represent, is fictive and entirely distinct from a physical molecule.) 

\textit{Derivation 2. Adiabatic Approximation from Heavy}. Recall that the position representation of $\hat T_1$ has the form $\partial^2/\partial x_1^2$:
\begin{equation}\label{eq:T1action}
    \hat T_1\theta_a\psi_a \propto \frac{\partial^2}{\partial x_1^2}\theta_a\psi_a=\frac{\partial^2\theta_a}{\partial x_1^2}\cdot\psi_a\ + 2\frac{\partial\theta_a}{\partial x_1}\cdot\frac{\partial\psi_a}{\partial x_1} + \theta_a\cdot\frac{\partial^2\psi_a}{\partial x_1^2}.
\end{equation}
However, $\theta_a\psi_a$ diagonalizes both $\hat T_2 +\hat W$ using (\ref{eq:RedHamTISE}), and (approximately)  $\hat H=\hat T_1+\hat T_2 +\hat W$ from (\ref{eq:BOansatz}). Therefore it also (approximately) diagonalizes $\hat T_1$: $\langle\theta_b\psi_b |\hat T_1|\theta_a\psi_a\rangle\propto\delta_{a,b}$. In the $x$-representation,
\begin{equation}\label{eq:BOexpand2}
 \int\mathrm{d}x_1\theta^*_b(x_1)\int\mathrm{d}x_2\psi^*_b(x_1,x_2)\ \Big\{\frac{\partial^2\theta_a}{\partial x_1^2}\psi_a\ + 2\frac{\partial\theta_a}{\partial x_1}\frac{\partial\psi_a}{\partial x_1} + \theta_a\frac{\partial^2\psi_a}{\partial x_1^2}\Big\}\propto\delta_{a,b}.
\end{equation}
Since different $\psi_a$ are orthogonal (since distinct eigenstates) the $x_2$ integral means that the first term in the sum is proportional to $\delta_{a,b}$. However, neither of the derivatives of $\psi_a$ will be orthogonal to $\psi_b$ (and similarly for the $\theta$s), so that the remaining terms will not be proportional to $\delta_{a,b}$ -- unless they are both zero. Thus (\ref{eq:BOexpand2}) entails that
\begin{equation}\label{eq:BOapprox}
    \frac{\partial\psi_a(x_1,x_2)}{\partial x_1}\approx0,
\end{equation}
(which is the more specific statement that often goes under the name the `Born-Oppenheimer approximation'). \begin{flushright}
$\square$
\end{flushright}

Note that \eqref{eq:BOapprox} is `adiabatic' in the sense that $\psi$ changes `slowly' with respect to $x_1$, not time (c.f. \cite{huggett:2023}).

Let us repeat, for emphasis, the important features of this mathematical discussion. First, although the ratio of electron to neutron mass is relevant, it only provides an inequality, without taking any limit (and in any case merely motivates an ansatz). Second, at no point does one approximate the molecule as having `clamped' nuclei; the role of (\ref{eq:RedHamTISE}) is to define a convenient basis in which an exact formal expansion can be made. As we have just seen, the crucial property of the basis is that -- because of the mass ratio -- one term in the expansion dominates for certain energy eigenstates of the molecule -- nothing more. Third, every step of the derivations is fully quantum. And finally, they suffice to show the mathematically validity of the \textbf{BO} approximation (in cases in which the \textit{Heavy} ansatz is correct).\footnote{Or, more precisely, and in line with various cautions that we have made, they sketch the more rigorous, equally quantum, derivation of the mathematical physicists.} 

That is, making the adiabatic approximation \eqref{eq:BOapprox} in \eqref{eq:T1action} yields
\begin{equation}\label{eq:T1approx}
    \hat T_1\theta_a\psi_a \approx \psi_a\frac{\partial^2\theta_a}{\partial x_1^2} = \psi_a\hat T_1\theta_a.
\end{equation}
In this specific sense, the part of the joint state $\Psi_a(x_1,x_2)$ that expresses the kinetic energy of the nuclei can (approximately) be factored out as $\theta_a(x_1)$. But to repeat the discussion after (\ref{eq:BOexpand1}), it cannot be over-emphasized that $\Psi_a(x_1,x_2)$ has \textit{not} been factored into strictly nuclear and electronic parts, since the other factor, $\psi_a(x_1,x_2)$, depends on both, not just the electron configuration.

Then, from (\ref{eq:BOansatz}), one needs to find $\theta_a(x_1)$ and $\psi_a(x_1,x_2)$; the latter is given by (\ref{eq:RedHamTISE}), so all we need is the equation for the former. From (\ref{eq:BOansatz}) we have: 
\begin{equation}
    \big(\hat T_1 + \hat T_2 + \hat W\big)\theta_a(x_1)\psi_a(x_1,x_2)
    \approx 
    E_a\theta_a(x_1)\psi_a(x_1,x_2),
\end{equation}
while from (\ref{eq:RedHamTISE}) and (\ref{eq:T1approx}) we have 
\begin{equation}
           \approx 
    \big(\hat T_1\theta_a(x_1)\big)\cdot\psi_a(x_1,x_2) + \lambda_a(x_1)\theta_a(x_1)\psi_a(x_1,x_2),
    \end{equation}
which gives:
\begin{equation}\label{eq:AdiabaticTheta}
    (\hat T_1 + \lambda_a(x_1)-E_a)\theta_a(x_1) \approx 0.
\end{equation}
This has the \textit{form} of a (time-independent) Schr\"odinger equation for the nuclear variables, `living' on a PES $\lambda_a(x_1)$, but recall the discussion after (\ref{eq:T1approx}). That cannot be the correct \textit{literal} description of the nuclei in \textbf{BO}, since \textit{both} $\theta_a(x_1)$ and $\psi_a(x_1,x_2)$ represent aspects of the nuclear subsystem. 

To sum up, from the Born-Oppenheimer separation ansatz  (\ref{eq:BOansatz}), finding the approximate eigenstates $\Psi(x_1,x_2)$ of $\hat H$ reduces to finding solutions to (\ref{eq:RedHamTISE}) and (\ref{eq:AdiabaticTheta}), and taking their product, a significant simplification. (Of course, these equations can still not be solved analytically, and require further approximations, for instance the WKB approximation.) \textit{Together} these steps constitute \textbf{BO}, and as we have shown, if \textit{Heavy }is true then they are fully quantum. Of course, as we have explained, making that ansatz is only valid if the results are (approximate) solutions in which \textit{Heavy} holds; since they are, \textbf{BO} is fully quantum. 

\subsection{Quantum Representation and the Uncertainty Principle}\label{sec:HUP}

In the light of the textbook treatment it should be apparent why the quoted claims about violations of Heisenberg uncertainty are inconsistent with the representational structure of \textbf{BO}. The crucial assumption  \textit{Heavy} involves no conflict with or suspension of the Heisenberg uncertainty principle. Furthermore, a violation of the Heisenberg uncertainty principle is formally precluded by the definition of the objects that are playing representational roles within \textbf{BO}.   

\textit{Heavy} is entirely consistent with the principles of quantum theory, including Heisenberg's. Indeed, the previous discussion makes clear that \textbf{BO} presupposes the standard wavefunction-observable representational structure of quantum theory, and hence \textit{entails} the uncertainty relations. First, wavefunctions are square-integrable functions on (configuration) space $L^2(\mathbb{R}^n)$. Second, (for a single degree of freedom) the variances of position and momentum variables are: $
 \sigma _{x}^{2}=\int _{-\infty }^{\infty }x^{2}\cdot |\psi (x)|^{2}\,dx\; \text{ and  }\;
\sigma _{p}^{2}=\int _{-\infty }^{\infty }p^{2}\cdot |\tilde{\psi} (p)|^{2}\,dp$, 
where $\psi (x)$ and $\tilde{\psi} (p)$ are the position and momentum basis wavefunctions, respectively. As is well-known, the Heisenberg uncertainty relation, $ \sigma _{x} \sigma _{p} \geq \frac{\hslash}{2}$, is then a simple mathematical consequence of the fact that position and momentum are Fourier conjugates.

A violation of the relation would require a `wavefunction' sharply peaked in \textit{both} the position and momentum basis simultaneously, contrary to this basic fact about QM and \textbf{BO}. They would only be violated if, the model represented electron or nuclei states as other than vectors in Hilbert space: say, as a delta functions in both position and momentum space.\footnote{See Footnote \ref{mixed} for a short discussion of precisely such a possibility in the context of mixed classical-quantum models in which there is failure of positivity of the density matrix in the quantum part of the model. As discussed there, this is understood by the scientists themselves as a pathological feature of the models rather than a putative representation of Heisenberg uncertainty violation.} Similarly, that the `nuclear wavefunction' $\theta_a(x_1)\in L^2(\mathbb{R}^m)$ is a vector in a Hilbert space means that it cannot violate the Heisenberg uncertainty relation; it does not and cannot violate the Heisenberg uncertainty relation. Moreover, as already noted, there is no justification for  interpreting  $\Psi_a(x_1,x_2)$ as factored into strictly nuclear and electronic parts. The factor, $\psi_a(x_1,x_2)$, depends on the nuclear configuration, not just the electron configuration (hence the scare quotes around `nuclear wavefunction'). The role of `classical' parameters and the `clamped' Hamiltonian in constructing $\psi_a(x_1,x_2)$ is not relevant to the status of the nuclei as quantum particles since all representation of states is via vectors in a Hilbert space

The crucial point is that \textbf{BO} provides an \textit{inherently quantum} representation of molecular structure that does not admit a classical separation into purely electronic and nuclear representations. The quantum nature of the nuclei in the \textbf{BO} is highlighted in the discussion of \cite{Jecko:2014} who notes in his concluding section: `We emphasise that, in the mathematical treatment of the Born-Oppenheimer approximation, the nuclei are always considered as quantum particles. The use of clamped nuclei is just a tool to construct an appropriate effective Hamiltonian but the latter is a quantum, nuclear Hamiltonian with restricted electronic degrees of freedom' \cite[p. 20]{Jecko:2014}. We will return to the quantum status of \textbf{BO} in the context of our more mathematically rigorous treatment in the next section.

\section{On the Mathematical Treatment of Born-Oppenheimer}
\label{Jecko}

Our explication of \textbf{BO} was, as we have explained, at a textbook level, eliding details and rigour for the sake of the larger mathematical picture. In this section we will discuss certain salient details, while still leaving others to the mathematical literature. In particular, there is an issue raised by \cite{sutcliffe2012quantum, Sut+Woo13} (and elsewhere in their work), critiquing \cite{Born:1954},  sometimes cited in support of the philosophical literature that we have criticized.

\cite{Sut+Woo13} point out that the part of the total molecular Hamiltonian (\ref{eq:HamTot}) containing the contributions from electronic kinetic energy and total Coulomb potential energy, $\hat T_2+\hat W$, which they denote $\hat H^{\mathrm{elec}}$, has a continuous spectrum (pp. 18-20). (Because, they argue, $\hat H^{\mathrm{elec}}$ is the direct integral of the $x_1$-parameterized family of clamped Hamiltonians, $\hat T_2+\hat W(x_1)$.) Of course, this fact is consistent with the discrete spectrum of the full Hamiltonian. However, it implies that (a) the PESs cannot be understood as representing the spectrum of $\hat H^{\mathrm{elec}}$, since they are discrete, and (b) \textbf{BO} cannot be understood as a \textit{perturbation} of $\hat H^{\mathrm{elec}}$, with small corrections due to nuclear motion: they could not make the spectrum discrete (also $\hat H^{\mathrm{elec}}$ gives no PESs for finding a nuclear perturbation). 

Sutcliffe and Woolley give as their target a common story about the \textbf{BO}, which holds the contrary view concerning these points, and their refutation is convincing. However, it does not appear to cause any difficulties for our explication. (a) The PESs are defined by the \textit{individual} members of the parameterized family of clamped Hamiltonians, not by their direct integral, $\hat H^{\mathrm{elec}}$; speaking extremely loosely, they are the `direct integral' of their eigenvalues, not the eigenvalues of their direct integral. (b) $\hat H^{\mathrm{elec}}$, as opposed (again) to the family of clamped Hamiltonians, plays no role in our exposition, which therefore certainly does not represent the \textbf{BO} as a perturbation about it! Indeed, \cite[p.28]{Sut+Woo13} seems to agree that a proper formulation of \textbf{BO}, in the kind of terms we have described allows `a rigorous account of the separation of electronic and nuclear motion'.

Regarding (a), note that the derivation itself shows why the PESs are relevant to the  \textbf{BO}: we have seen that they are precisely the functions that appear, as a matter of mathematics, as potential terms in (\ref{eq:AdiabaticTheta}). No more explanation is needed, and no need to think of any literal clamping occurring. Of course the role of the PES is much richer than this in quantum chemistry, and far more can be said about its general physical significance (something we touch on in \S\ref{BBO}). Again, this paper is focussed specifically on \textbf{BO}, not on chemistry in general, though of course we postulate that the kind of analysis presented can be generalized.

\cite{Jecko:2014} invokes Sutcliffe and Woolley's arguments and `slightly pessimistic note', to review the current state of mathematical knowledge regarding \textbf{BO} and, specifically, to explain, more optimistically, how mathematical physicists have dealt with a distinct issue regarding continuous spectra.\footnote{In personal communications, both Jecko and Woolley have said that (a) the Born-Huang approach is formally flawed (as argued in \cite{Sut+Woo13}), (b) the more rigorous approach found in \cite{Jecko:2014} avoids those flaws, and (c) extant rigorous treatments do not currently capture every important chemical concept. Jecko's paper aimed to clarify (b) and some successes in the direction of (c). Neither party claims that chemistry can or cannot be reduced to physics, though they may assess the prospects somewhat differently.} The problem is that the `clamped' Hamiltonians used in \textbf{BO} generally have a continuous part to their spectra (representing energies at which the constituents have disassociated, and there is no longer a stable molecule), above the discrete part (corresponding to a bound state). Our analysis assumed that the spectra are purely discrete -- the existence of a PES depends on it -- so it is crucial to our interpretive claims to understand the nature and justification of this assumption. Indeed, Jecko's discussion identifies three principal problems for making the textbook treatment of \textbf{BO} exact, each of which we should address.\footnote{Thierry Jecko has also emphasized to us in correspondence that there are rigorous approaches to the quantum treatment distinct from \textbf{BO}, for instance the `exact factorization' of \cite{grossexactfac}, discussed in \cite{jecko2015factorization}.} 

First, the analysis only involves the internal energy of the molecule, even though it has centre of mass motion and associated kinetic energy, with a continuous spectrum (absent boundary conditions). To separate out the internal levels (to find the spectrum, or the spatial structure, or for scattering) one can transform to a centre of mass frame (of the molecule or just its nuclei, depending on the problem), and subtract the centre of mass kinetic energy. This procedure introduces a new term into the Hamiltonian (the Hughes-Eckart energy), which is suppressed by the ratio of electronic to nuclear masses ($\sim10^{-3}$), so is neglected in low order approximation.\footnote{\cite{sutcliffe2012quantum} caution that ignoring it can make the Hamiltonian ill-defined.} This complication is harmless for present purposes, so we suppressed it for reasons of space.

The second problem arises because the expansion (\ref{eq:BOexpand1}) is not really a sum, but a sum of low energy states plus an \textit{integral} over states in the high energy, continuous part of the spectrum of $\hat T_2+\hat W$. As is familiar, in any continuous spectrum there are no normalizable eigenstates, something finessed in familiar ways by physicists via the Dirac delta function, and more rigorously by the theory of `spectral decomposition'. However, in the present case details of the structure of the spectrum, in particular the existence of `thresholds', \cite[\S V]{Jecko:2014},  make the expansion highly non-trivial, and hard to control mathematically, due to the possibility of divergences at the thresholds. Its existence is thus -- from a mathematically rigorous point of view -- an `in principle' matter only. 

The third `problem' is that the appropriate formalism for a mathematically rigorous treatment is, as discussed earlier, that of the \textit{direct integral} \cite[280-7]{reed1978iv}. Any $L^2(\mathbb{R}^m\times\mathbb{R}^n)$ function $f(x_1,x_2)$ is understood as an $L^2(\mathbb{R}^n)$-valued function $f(x_1,\cdot)$ with $x_1\in\mathbb{R}^m$. One then naturally defines the direct integral of a parameterized family of operators $\hat O(x_1)$ on $L^2(\mathbb{R}^n)$, as the operator $\hat O$ on $L^2(\mathbb{R}^m\times\mathbb{R}^n)$ whose effect on $f(x_1,x_2)$ is the direct integral of $\hat O(x_1)f(x_1,\cdot)$: i.e., it acts on each $x_2$ function as the appropriate operator in its state space.\footnote{There are important questions of the self-adjointness of the various Hamiltonians involved. And we continue to ignore degeneracy in the spectrum.} So one needs to be sure that the formalism is respected.

In response to the last two problems, mathematical physicists have a somewhat different perspective on \textbf{BO} to that of \S\ref{textbook}.\footnote{The history of the development of more rigorous approaches to \textbf{BO} is complex and fascinating. Initial steps were undertaken in \cite{combes:1975,Combes:1981}. See \cite{sutcliffe2012quantum} and \cite{Jecko:2014} for more details.} The goal is to approximate an eigenstate (and eigenvalue) lying in the low energy, discrete part of the spectrum of  the molecule (\ref{eq:HamTot}). To do this, one considers eigenstates of the Hamiltonian \textit{projected} onto the subspace in which its spectrum is discrete. See  \cite[\S IV]{Jecko:2014} for details of the following sketch. 

As noted, $\hat T_2+\hat W$ is the direct integral of the clamped Hamiltonians $\hat T_2+\hat W(x_1)$. While the latter act on $L^2(\mathbb{R}^n)$, the state space of the electron subsystem, the former acts on $L^2(\mathbb{R}^m\times\mathbb{R}^n)$, the state space of the full system, including both the electrons \textit{and the nuclei}: both, that is, are quantum mechanical. $\hat T_2+\hat W$ is also often called the `electronic Hamiltonian' (e.g., $\hat H^{\mathrm{elec}}$ above), but that would be misleading for current purposes since it acts on the state of the \textit{whole} molecule. Similarly for projection operators: for a given $x_1=X$ and eigenstates $\psi_a(X; x_2)$ of $\hat T_2+\hat W(X)$ lying in a \textit{finite discrete} part of its spectrum there is an operator 
\begin{equation}
    \hat P(X)f(x_2)=\sum_{a=1}^N\psi_a(X; x_2)\int\psi^*_a(X; x'_2)f(x'_2)\text{d}x'_2\equiv\sum_{a=1}^N\phi_a(X)\cdot\psi_a(X; x_2)
\end{equation}
projecting $f(x_2)\in L^2(\mathbb{R}^n)$ onto the subspace spanned by the $\psi_a(X; x_2)$. Then, $\hat P$, the direct integral of the $\hat P(x_1)$ projects states in $L^2(\mathbb{R}^m\times\mathbb{R}^n)$ onto the proper subspace, $V$, spanned by the direct products of the $\phi(X)\psi_a(X; x_2)$. Since $\hat P$ also acts on the full molecular states space, it too treats both nuclei and electrons as quantum mechanical. 

To overcome the issue of the continuous part of the clamped Hamiltonian's spectrum, the approximation made (see   \cite[\S IV]{Jecko:2014} for justification) is to seek eigenstates for the \textit{projected} molecular Hamiltonian $\hat H_{\text{eff}}=\hat P(\hat T_1+\hat T_2+\hat W)\hat P$, instead of  for $\hat T_1+\hat T_2+\hat W$. One assumes that if an eigenstate of the latter exists in a given narrow range of the energy, then it can be approximated by an eigenstate of the former lying in the same range: \textit{one replaces the tricky problem of solving the `true' Hamiltonian with the simpler problem of solving an effective Hamiltonian for the energy range of a stable molecule}. Formally, $\hat H_{\text{eff{}}}$ acts on $L^2(\mathbb{R}^m\times\mathbb{R}^n)$: first a state is projected onto $V$, then acted on by the molecular Hamiltonian, and the result projected again onto the subspace. $\hat H_{\text{eff{}}}$ too treats both nuclei and electrons as quantum. Note that the effect of $\hat H_{\text{eff{}}}$ is always to produce a vector in $V$, so while (a) all states orthogonal to $V$ are eigenstates of $\hat H_{\text{eff{}}}$ with eigenvalue 0, (b) all its other eigenstates lie in $V$.\footnote{From its form, one expects at least a discrete part of the spectrum, containing the bound states of interest.} The upshot is, in relation to \S\ref{textbook}-\ref{sec:derivations}, that by excising the continuous part of the spectra, one has made rigorous (\ref{eq:BOexpand1}), now understood as expanding solutions of the Schr\"odinger equation for the \textit{projected} Hamiltonian, within the defined energy range.

The previous sections show that textbook \textbf{BO} is fully quantum, and adding a step in which the Hamiltonian is projected onto a subspace of the Hilbert space in no way undermines that argument. The projected Hamiltonian is still quantum, indeed a quantum theoretical approximation to the quantum Hamiltonian. The difference is rather that \textbf{BO} is no longer understood as the lowest order in some well-defined exact expansion, but rather as an approximation, with some well-understood corrections, to an exact solution. This situation is in contrast, not only to our textbook style presentation, but also to those of Born and Oppenheimer, Born and Huang, and Messiah; while more careful than ours, theirs remain heuristic.\footnote{See \cite[\S V]{Jecko:2014} for a detailed comparison.} Once the appropriate formal machinery has been deployed, \textbf{BO} can be formalised to the standards of rigour of mathematical physics.  No assumptions inconsistent with quantum theory are involved. 

\section{Rigour and Reduction in Quantum Chemistry}
\label{BBO}

This final section considers questions of mathematical rigour and reduction in quantum chemistry more generally. Let us start with the role of `classical' assumptions in the emergence of `determinate molecular structure'. Let us first assume that what is meant by `determinate molecular structure' is provided by the regime in which the adiabatic approximation is valid, and thus the nuclei are such that the variation of the `electron' wavefunction with respect to nuclear positions is approximately zero. \S\ref{sec:BOapprox}-\ref{Jecko} show exhaustively that the emergence of such structure requires neither classical assumptions in the sense of clamped nuclei, nor violation of the uncertainty relation. Rather the assumption required for \textbf{BO} is \textit{Heavy}, which is entirely consistent with a fully quantum treatment of the molecule. There is a richer notion of molecular structure that includes the number, angles and lengths of chemical bonds and chirality \citep{franklin:2020}. \textbf{BO} is far from sufficient to model such structure, although plausibly in some circumstances it may prove necessary.\footnote{It is certainly not the case that the \textbf{BO} is necessary in general. See Footnote \ref{LBBO} for references on quantum chemistry \textit{beyond} the \textbf{BO}.} In this context, there is debate about the requirement for a solution to the measurement problem of quantum theory in order to account for the emergence of molecular structure \citep{franklin:2020,fortin:2021,seifert:2022,miller:2023}. However, it remains to be seen which foundational features, if any, mark out a \textit{particular} distinctiveness of quantum chemistry, within the vast array of applications of quantum mechanics to diverse matter systems, including non-molecular states of matter such as plasmas or Bose-Einstein condensates.

More generally, in the context of discussions of `classical' and `semi-classical' assumptions our positive proposal for precise scientific language use is as follows. Firstly, models such as \textbf{BO} should be described as `quantum' models of molecules, since all aspects of the molecules are treated quantum mechanically within the Hilbert space formalism. This is not withstanding the fact that \textbf{BO} can be understood as `mimicking' a semiclassical Schr\"odinger operator due to the functional analogy between the the nuclear kinetic energy term and a semiclassical Laplace operator \cite[p.19]{Jecko:2014}. Furthermore, on this proposed usage adiabaticity is neither necessary or sufficient for semi-classicality. Quantum models can be adiabatic and semi-classical models can be non-adiabatic.

Secondly, one may legitimately deploy the term `semi-classical' in a broad sense to indicate an application of quantum theory in which, while the system is treated quantum mechanically, the model includes radiative fields that are not. In this sense, the original quantum theory of the atom, as well as the modern quantum-mechanical treatment, is `semi-classical' since the electromagnetic field through which the nucleus and the electrons interact is not quantized, cf. \citep{boucher:1988}. Similarly for most of quantum chemistry including \textbf{BO}. It would be moot to claim that chemistry does not reduce to quantum physics because the physics in question is semi-classical in this sense, because huge amounts of what is ordinarily called `quantum physics' is! (For example, standard non-relativistic quantum-mechanics other than free particles, including most condensed matter physics.) Such semi-classical physics is a kind of quantum physics. 

Thirdly, one should distinguish various more specific meanings of `semi-classical'. One refers to a quantum mechanical expansion, in terms of classical zeroth order plus quantum corrections, which is truncated at some order in $\hslash$ (or other parameter such as mass ratios) to provide an approximate model.\footnote{The two most important examples are when such a truncation is made in a quantum moment expansion (typically leading to Ehrenfest type equations) or in an expansion for the wavefunction (typically leading to a WKB-approximation).} In such cases, the semi-classical model is a sub-model within the general framework of quantum theory.\footnote{We thus recover the idea familiar from Nickles reduction of a successor theory containing a version of the predecessor theory via the application of a set of mathematical operations to its models \citep{palacios:2022}.} A formally closely related, but physically and conceptually different, sense of `semi-classical' is found in the context of certain `semi-classical' approaches to gravity where the field equations are re-written in terms of classical metric variables on the left-hand side but the first moment (i.e. expectation value) of the stress-energy tensor is inserted on the right-hand side. A further notion of semi-classicality refers to specific behaviour in the limit (usually $\hslash\rightarrow0$) in which the classical theory obtains.\footnote{Such cases have been much discussed in the physics and philosophy of physics literature  \citep{,berry:1977,berry2001chaos,batterman:2001,bokulich:2008,rosaler:2015,steeger:2021,feintzeig:2020,DawidThebault2025} and they merit study in the context of quantum chemistry.}  Finally, the term `semi-classical' is applied to mixed classical-quantum models which simultaneously feature representations of classical and quantum states and dynamics. The former in terms of classical phase space states and Poisson bracket structure and the latter in terms of operators, density matrices and commutator brackets structure. Mixed classical-quantum models have been widely applied in non-adiabatic quantum chemical modelling \citep{tully:1991,crespo2018recent} and bring up various interesting issues that are worthy of philosophical engagement. 

It is useful to distinguish the distinct modelling contexts in which semi-classical models, in one or more of the more specific senses, might be deployed within quantum chemistry and articulate the relevance for reductive explanations in each context. They are as follows. (1) \textit{Mathematical Idealization}: The model is being used to represent a fully quantum phenomena and the semi-classical features are a mathematical idealization that relates the semi-classical effective model to a (less) idealized fully quantum model. Given the stability of salient explanatory features under de-idealization there no potential problem for reductive explanations of the relevant phenomena. (2) \textit{Physical Idealization}: The model is being used to represent a fully quantum phenomena but this representation is via proxy model of semi-classical phenomena that approximate the quantum phenomena for the purposes and degree of accuracy required. Again, there is no problem for reductive explanations given stability of the salient explanatory features under de-idealization \citep{bokulich:2008,bokulich:2017}. (3) \textit{Emergent Phenomena}: The model is being used to represent emergent semi-classical phenomena that occur in the context of classical-quantum limit behaviour. There is a putative problem for any account of reductive explanation that is incompatible with emergence qua novel and robust behaviour, but not for any account of reductive explanation that is so compatible, cf. \citep{bokulich:2008,butterfield:2011,franklin:2024}.

A further issue is the role of the environment \citep{ladyman:2024}. In this context, \cite{sutcliffe2012quantum} say `one should not expect useful contact between the quantum theory of an isolated molecule and a quantum account of individual molecules, as met in ordinary chemical situations where persistent interactions (due to the quantized electromagnetic field, other molecules in bulk media) and finite temperatures are the norm.' (p. 7). These remarks, together with related ideas developed by \cite{seifert:2022}, form a fruitful basis for an `open systems view' of quantum chemistry in the manner recently proposed for quantum physics more generally by \cite{cuffaro:2021}. We expect that such a view would prove to be consistent with any suitably nuanced understanding of model-based understanding of reduction and idealization.\footnote{Whilst it is common in quantum chemistry to model molecules as if they were isolated from their environment, real molecules interact with other systems as per the quote from \cite{sutcliffe2012quantum}. However, in models of an `isolated' quantum system the effects of the environment may be added in to produce an `autonomous' open system model, just as the effect of friction can be represented via an autonomous model of an `isolated' oscillator \cite{ladyman:2024}. We may thus, following \cite{seifert:2022}, understand the use of non-Coulombic Hamiltonians in quantum chemistry as a means of encoding persistent environmental interactions in a manner analogous to the use of non-conservative forces in other physical modelling contexts. The structure of such `open quantum chemistry' models is closely related to physical models of environmental decoherence. The connection between emergence-reduction in the context of decoherence \citep{wallace:2012,joos:2013,dawid:2015,franklin:2024,DawidThebault2025} and `open quantum chemistry' deserves further study.}

We conclude with the following questions: Is there a methodological distinction between the use of models in modern quantum chemistry, and other examples of quantum modelling practice in matter systems, such as applications of quantum mechanics to solid state or few body systems? Is there space for distinctively `chemical' modes of quantum modelling practice? In answering such questions we suggest that philosophers need to re-conceptualise the physics-chemistry interface and move beyond the confines of the reduction vs. anti-reduction dialectic. Plausibly, the characteristically `chemical' aspects of quantum chemistry are rooted in methodology rather than disparate formal features of the models.\footnote{In general, one would can expect methodological distinctiveness to be ubiquitous within the application of scientific models to matter systems in the domain of any `mid-range theory' \citep{cartwright:2020} such as, for example, hydrodynamics cf. \cite{batterman:2021}.}  For example, experimental practices, such as spectroscopy. On our view, we should understand quantum chemistry as neither a putative case of inconsistency between chemistry and physics nor an instance of subsumption of the former by the latter. The field of quantum chemistry should not be understood to mark a boundary between the distinctly chemical and distinctly physical nor, moreover, between classical and quantum modes in the representations of molecules.\footnote{In this regard our understanding is very much in keeping with the analysis of \cite{ramsey:1997}.} Rather, quantum chemistry should be recast as a `littoral zone' with its own distinctive modelling ecology, conditioned by its role as the fertile meeting place of the two disciplines.

\section*{Acknowledgements}

Thanks to Hern\'an Accorinti, Guido Bacciagaluppi, Hasok Chang, Sebastian Fortin, Robin Hendry, Juan Camilo Mart\'inez Gonz\'alez, Olimpia Lombardi, Vanessa Seifert, Guy Woolley, two anonymous referees, and audiences at several venues for responses to and discussions of drafts, which helped us sharpen our arguments, and avoid some errors. Thanks especially to Thierry Jecko for generous discussion and feedback.

\section*{Funding Statement}
This work was supported by the Benjamin Meaker Fellowship programme at the University of Bristol and by John Templeton Foundation grant number 62210.

\section*{Declarations}

None

\bibliographystyle{chicago}
\bibliography{BO}
\end{document}